\documentclass[%
aip,
amsmath,amssymb,
reprint,
]{revtex4-1}

\usepackage{graphicx}
\usepackage{dcolumn}
\usepackage{bm}
\usepackage[utf8]{inputenc}
\usepackage[T1]{fontenc}
\usepackage{mathptmx}
\usepackage{etoolbox}
\usepackage{physics}
\usepackage{bbding}
\usepackage{xcolor}
\DeclareMathAlphabet\mathbfcal{OMS}{cmsy}{b}{n}
\usepackage[cal=boondoxo]{mathalpha}

\begin{document}
\preprint{AIP/123-QED}

\title{Quantum Geometric Origin of Orbital Magnetization}

\author{Xiao-Bin Qiang}
\altaffiliation{These authors contributed equally to this work.}
\affiliation{State Key Laboratory of Quantum Functional Materials, Department of Physics, and Guangdong Basic Research Center of Excellence for Quantum Science, Southern University of Science and Technology (SUSTech), Shenzhen 518055, China}
\affiliation{Division of Physics and Applied Physics, School of Physical and Mathematical Sciences, Nanyang Technological University, 21 Nanyang Link, 637371, Singapore}

\author{Tianyu Liu} 
\altaffiliation{These authors contributed equally to this work.}
\affiliation{Department of Physics, Southern University of Science and Technology (SUSTech), Shenzhen 518055, China}

\author{Hai-Zhou Lu}
\email{luhz@sustech.edu.cn}
\affiliation{State Key Laboratory of Quantum Functional Materials, Department of Physics, and Guangdong Basic Research Center of Excellence for Quantum Science, Southern University of Science and Technology (SUSTech), Shenzhen 518055, China}
\affiliation{Quantum Science Center of Guangdong-Hong Kong-Macao Greater Bay Area (Guangdong), Shenzhen 518045, China}

\author{X. C. Xie}
\affiliation{International Center for Quantum Materials, School of Physics, Peking University, Beijing 100871, China}
\affiliation{Interdisciplinary Center for Theoretical Physics and Information Sciences (ICTPIS), Fudan University, Shanghai 200433, China}
\affiliation{Hefei National Laboratory, Hefei 230088, China}

\date{\today}

\begin{abstract}
The exploration of the Riemannian structure of the Hilbert space has led to the concept of quantum geometry, comprising geometric quantities exemplified by Berry curvature and quantum metric. While this framework has profoundly advanced the understanding of various electronic phenomena, its potential for illuminating magnetic phenomena has remained less explored. In this Perspective, we highlight how quantum geometry paves a new way for understanding magnetization within a single-particle framework. We first elucidate the geometric origin of equilibrium magnetization in the modern theory of magnetization, then discuss the role of quantum geometry in kinetic magnetization, and finally outline promising future directions at the frontier of quantum geometric magnetization.
\end{abstract}

\maketitle

\section{Introduction} 
Quantum geometry has emerged as a universal language, bridging the microscopic world of wave functions and the macroscopic world of physical phenomena~\cite{Provost80cmp, AA90prl, Resta11epjb, Berry84rspa, XiaoD10rmp}. This framework centers on Berry curvature and quantum metric as its fundamental components, underpinning a broad spectrum of manifestations in condensed matter systems~\cite{Torma23prl, Lu24nsr, YanBH25rppp}. In flat-band superconductors, the superfluid weight is governed by the Brillouin-zone integral of  quantum metric~\cite{Torma15nc, Torma16prl, Torma17prb, Torma18prb, Torma22prb}. In fractional Chern insulators, the stability of the topological phase is crucially based on nearly uniform distributions of Berry curvature and quantum metric throughout the Brillouin zone~\cite{Roy12prb, Roy14prb, Roy15nc}. In the quantum anomalous Hall effect, the Chern number is explicitly defined by the integral of the Berry curvature of the occupied bands~\cite{Haldane88prl, Lu10prb, YuR10science, XueQK13science, ChangCZ23rmp}. In nonlinear transport~\cite{KangKF19nm, XuSY19nature, Tiwari21nc, Kumar21nn, MaT22nc, KTLaw23nsr, XuSY23science, GaoWB23nature}, the nonlinear electrical conductivity can be determined by the dipole of Berry curvature or quantum metric~\cite{GaoY14prl, FuL15prl, Lu18prl, Lu21nc, YangSY21prl, GaoY21prl, YanBH24prl, Lu25as}.

The transformative role of quantum geometry in understanding electronic phenomena naturally leads to a fundamental question: Can this geometric perspective provide an equally powerful framework for magnetic phenomena? Magnetization, one of the oldest and most ubiquitous phenomena, has captivated scientific curiosity for millennia, but its microscopic origin has constituted a long-standing challenge in condensed matter physics~\cite{Ashcroft76book, Stephen01book}. Fortunately, quantum geometry provides the essential language to bridge this gap at a single-particle level, offering a theoretical framework connecting magnetic moments to various geometric quantities. In this Perspective, we first revisit the modern theory of magnetization to demonstrate its geometric foundations. We then highlight recent progress, with special emphasis on kinetic magnetization and its geometric interpretation. Finally, we provide an outlook on the emerging opportunities and open questions in the intriguing field of magnetization.

\section{Modern theory of magnetization}
Magnetization in general originates from two intrinsic degrees of freedom: spin and orbital.

In isolated atoms or molecules, magnetization (i.e., magnetic moment per unit volume) cannot be well defined, but both the spin and orbital magnetic moments (respectively denoted as $\mathbf{m}^{\text{S}}$ and $\mathbf{m}^{\text{O}}$) can be expressed in terms of the spin and orbital angular momenta (respectively denoted as $\mathbf S$ and $\mathbf L$) as $\mathbf{m}^{\text{S}}=-\mu_B\langle\mathbf S/\hbar\rangle$ and $\mathbf{m}^{\text{O}}=-\mu_B\langle\mathbf{L}/\hbar\rangle$ ~\cite{Griffith18book, Kahn21book}, where $\mu_B$ is the Bohr magneton and $\hbar$ is the reduced Planck constant. 

In crystalline materials, the definition of spin and orbital magnetization becomes more subtle, as electrons are characterized by Bloch waves, and correlation effects can start to manifest~\cite{Auerbach12book,NgTK17rmp,Vojta18rpp}. In this perspective, we restrict our scope to a single-particle framework. The spin magnetization now requires summing over all occupied electronic states and reads
\begin{equation}\label{Eq: M_spin}
\mathbf{M}^{\text{S}}=-\frac{\mu_B}{\hbar \mathcal{V}}\sum_{n,\mathbf{k}}\bra{n\mathbf{k}}  \mathbf S \ket{n\mathbf{k}}f_0,
\end{equation}
where $\mathcal{V}$ is the volume of the material under consideration, $\ket{n\mathbf{k}}$ denotes the unit-cell-periodic state with band index $n$ and crystal momentum $\mathbf{k}$, and $f_0\equiv f_0(\varepsilon_{n\mathbf{k}})$ is the Fermi-Dirac distribution function evaluated at the band energy $\varepsilon_{n\mathbf{k}}$. On the other hand, due to the quenching of the intra-atomic orbital magnetic moment by the crystal field~\cite{Bruno89prb,Marcelo96rmp,Abragam12book}, a major challenge arises when searching for orbital magnetization, because the position operator $\mathbf r$ cannot be well defined, which in turn makes the orbital angular momentum and thus the orbital magnetic moment ill-defined. This conceptual difficulty persisted for nearly a century until it was finally resolved in recent decades.

Two complementary approaches~\cite{Niu96prb, Niu99prb, Niu05prl, Resta05prl, Resta06prb, Niu07prl} have been developed to overcome this difficulty. On the one hand, the problem can be reformulated with Wannier functions, whose strong real-space localization naturally leads to a well-defined position operator~\cite{Resta05prl, Resta06prb}. On the other hand, Bloch waves can be reorganized into a wave packet, whose center helps regularize the definition of position operator and thus enables the calculation of various physical observables~\cite{Niu96prb, Niu99prb, Niu05prl, Niu07prl}. As the two approaches are in essence equivalent, we choose to illustrate the modern understanding of orbital magnetization with the wave packet formulation. 

\begin{figure}[t]
\centering 
\includegraphics[width=0.25\textwidth]{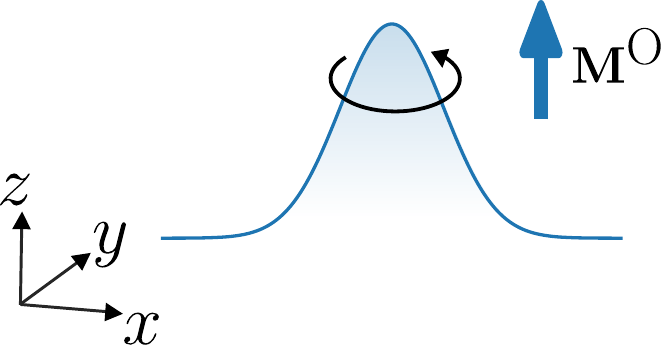}
\caption{
Wave packet self-rotation as the microscopic origin of the orbital magnetic moment. The black and blue arrows indicate the self-rotation and the resulting magnetization, respectively.
}
\label{Fig: equilibrium}
\end{figure}

The wave packet of the $n$th band can be constructed from the corresponding Bloch states $\ket{\psi_{n\mathbf{k}}}=\mathrm e^{\mathrm{i}\mathbf{k}\cdot\mathbf{r}}\ket{n\mathbf{k}}$ as~\cite{Niu96prb, Niu99prb, XiaoD10rmp}
\begin{equation}
\ket{W_n}=\int d\mathbf{k}w(\mathbf{k})\ket{\psi_{n\mathbf{k}}},
\end{equation}
where the weight $w(\mathbf{k})$ satisfies $|w(\mathbf{k})|^2 \simeq \delta(\mathbf{k} - \mathbf{k}_c)$ with $\mathbf{k}_c$ labeling the central momentum of the wave packet. Within this formulation, the expectation of the position operator $\mathbf{r}_c = \bra{W_n}\mathbf{r}\ket{W_n}$ converges, marking the positional center of the $n$th wave packet. The orbital magnetic moment for the $n$th band $\mathbf{m}_n^{\text{O}}=-(e/2)\bra{W_n}(\mathbf{r}-\mathbf{r}_c)\times\dot{\mathbf{r}}\ket{W_n}$ with electron charge $-e$ then becomes well-defined, and can be pictorially understood as the self-rotation of the wave packet~\cite{XiaoD10rmp,Dimi24apx}(Fig.~\ref{Fig: equilibrium}). In a more practical way, the orbital magnetic moment can be rewritten as
\begin{equation}\label{Eq: m_n_def}
\mathbf{m}_n^{\text{O}}=\frac{e}{2\hbar}\text{Im}\bra{\partial_\mathbf{k}n}\times(\mathcal{H}-\varepsilon_n)\ket{\partial_\mathbf{k}n},
\end{equation}
where $\mathcal{H}$ is the Hamiltonian of the material under consideration, and $\varepsilon_n\equiv\varepsilon_{n\mathbf k}$ and $\ket{n}\equiv \ket{n\mathbf k}$ are adopted for transparency.


To better understand the geometric origin of the orbital magnetic moment, it would be inspiring to compare the orbital magnetic moment tensor with the Berry curvature tensor. Their expressions are~\cite{XiaoD10rmp}
\begin{equation} \label{Eq: mO}
\left\{
\begin{aligned}
&m_n^{\text{O},ij}=-\frac{e}{\hbar}\text{Im}\sum_{m\neq n} \frac{\langle  n|\partial_i \mathcal{H}|m\rangle\langle m| \partial_j \mathcal{H}|n \rangle}{\varepsilon_n-\varepsilon_m},
\\
&\Omega_n^{ij}=-2\text{Im}\sum_{m\neq n}\frac{\langle  n|\partial_i \mathcal{H}|m\rangle\langle m| \partial_j \mathcal{H}|n \rangle}{(\varepsilon_n-\varepsilon_m)^2},
\end{aligned}
\right.	
\end{equation}
which are respectively related to their vector forms through $m_n^{\text{O},k}=\epsilon_{ijk} m_n^{\text{O},ij}$ and $\Omega_n^{k}=\epsilon_{ijk} \Omega_n^{ij}$ with $\epsilon_{ijk}$ being the Levi-Civita symbol. It is worth noting that the orbital magnetic moment tensor only differs from the Berry curvature tensor by a coefficient and a band normalization. The formal similarity of their expressions implies an intrinsic quantum geometric nature of the orbital magnetic moment.

As a concrete example, we consider the two-dimensional massive Dirac model~\cite{Shen17book}, whose Hamiltonian is given by $\mathcal{H}=v(k_x\sigma_x+k_y\sigma_y)+m\sigma_z$, with $v$ and $m$ being the model parameters. According to Eq.~\eqref{Eq: mO}, the Berry curvature of the conduction band reads~\cite{YaoW07prl, Lu15prb, Lu23prb, Lu24nsr} $\Omega_+^{z}=-mv^2/(2d^3)$ with $d=(v^2k^2+m^2)^{1/2}$, while the orbital magnetic moment reads $m_+^{\text{O},z}=-e m v^2/(2\hbar d^2)=e d\Omega_+^{z}/\hbar$, implying that the orbital magnetic moment is an inherently geometric quantity.

The quantum geometric origin of the orbital magnetic moment strongly suggests a profound geometric nature of the orbital magnetization, whose expression reads~\cite{Niu05prl}
\begin{equation}\label{Eq: M_orbi}
\mathbf{M}^{\text{O}}=\frac{1}{\mathcal{V}}\sum_{n,\mathbf{k}}\mathbf{m}_n^{\text{O}}f_0 - \frac{e}{\hbar\beta\mathcal{V}}\sum_{n,\mathbf{k}}\boldsymbol{\Omega}_n\ln (1-f_0).
\end{equation}
The first term of Eq.~\eqref{Eq: M_orbi} is a summation of the orbital magnetic moment $\mathbf{m}_n^{\text{O}}$ weighted by the distribution function $f_0$, analogous to the expression of the spin magnetization [Eq.~\eqref{Eq: M_spin}]. More profoundly, an additional geometric correction emerges. These two contributions form the foundation of the modern theory of orbital magnetization~\cite{Niu05prl, Resta05prl, Resta06prb, Niu07prl, Krister19jpcs}, highlighting the quantum geometric nature of orbital magnetization.

\section{Kinetic Magnetization}

In time-reversal-symmetric systems, equilibrium spin and orbital magnetization [Eqs.~\eqref{Eq: M_spin} and~\eqref{Eq: M_orbi}] vanish identically. However, an external electric field \(\mathbf{E}\) can drive the system out of equilibrium, giving rise to a correction in the distribution function~\cite{Hurd72book,Haug08book}, $f_1=e\tau\mathbf{v}_n\cdot\mathbf{E}f_0'$, where $\tau$ is the relaxation time, $\mathbf{v}_n=\partial_\mathbf{k}\varepsilon_n/\hbar$ is the group velocity, and $f_0'\equiv\partial f_0/(\partial\varepsilon_n)$. Such a nonequilibrium correction then induces a net magnetization known as the extrinsic kinetic magnetization. Furthermore, in systems where time-reversal symmetry is broken, the electric field can directly couple to quantum geometric properties of the electronic states, giving rise to an intrinsic kinetic magnetization. Both effects can be generally expressed as $\delta\mathbf{M}=\boldsymbol{\alpha}\mathbf{E}$, with $\boldsymbol{\alpha}$ being the response tensor.

\subsection{Spin contribution}
In the presence of time-reversal symmetry, the electric-field-induced kinetic spin magnetization can be readily evaluated by replacing $f_0$ with $f_1$ in Eq.~\eqref{Eq: M_spin}. Explicitly, it reads
\begin{equation} \label{Eq: delta_Ms_ex}
\delta\mathbf{M}^{\text{S},ex}=-\frac{e\tau\mu_B}{\hbar \mathcal{V}}\sum_{n,\mathbf{k}}\bra{n} \mathbf S \ket{n} \mathbf{v}_n\cdot\mathbf{E}f_0'.
\end{equation}
This prominent example was proposed by Edelstein~\cite{Edestein90ssc}, who showed that an applied electric field generates spin magnetization/accumulation in systems with strong spin-orbit coupling. This phenomenon, now widely known as the spin Edelstein effect, plays an indispensable role in spintronics~\cite{Manchon19rmp}.

We note that the above discussion is formulated within the relaxation-time approximation, where extrinsic contributions are treated at the level of a constant relaxation time $\tau$. Beyond this simplified framework, more intricate scattering mechanisms, such as skew scattering and side jump, are known to play essential roles in the anomalous Hall effect~\cite{Nagaosa10rmp} and the spin Hall effect~\cite{Sinova15rmp}. These mechanisms may give rise to additional contributions for kinetic spin magnetization. A recent theoretical study~\cite{XiaoC24prb} has explored such possibilities, pointing to a richer landscape of scattering-induced spin magnetization.

When time-reversal symmetry is further broken, an additional kinetic spin magnetization emerges geometrically as~\cite{Franz10prl, Dimi17prb, ChenYY24prb, Johansson24jpcm}
\begin{equation}\label{Eq: delta_Ms_in}
\delta\mathbf{M}^{\text{S},in}=-\frac{e\mu_B}{\hbar\mathcal{V}}\sum_{n,\mathbf{k}}\boldsymbol{\Phi}_n\cdot \mathbf{E}f_0,
\end{equation}
where $\boldsymbol{\Phi}_n$ is the spin Berry curvature tensor with components $\Phi_n^{ij}=-2\text{Im}\sum_{m \neq n}\langle n|S_i|m \rangle \langle m|\partial_j\mathcal{H}|n \rangle/(\varepsilon_n -\varepsilon_m)^2$. In contrast to that arising from the Edelstein effect, this kinetic spin magnetization is purely intrinsic (i.e., independent of $\tau$). 

The kinetic spin magnetization arising from the Edelstein effect and the geometric mechanism has been experimentally observed in a variety of materials, including $\beta$-Ta~\cite{Liu12science}, (Ga,Mn)As~\cite{Sinova14nn}, Bi$_2$Se$_3$~\cite{Jonker14nn}, $\alpha$-Sn~\cite{Fert16prl}, and Mn$_3$Sn~\cite{Otani19nature}. It is important to note that in many of these heavy metals and topological materials, the bulk kinetic spin magnetization often coexists with the boundary spin accumulation arising from the spin Hall effect~\cite{Perel71pla,Hirsch99prl,Kato04prl,Sinova15rmp}. Disentangling these two contributions, bulk magnetization versus boundary accumulation, remains a subtle but critical aspect in both experiment and theory.

\begin{figure}[t]
\centering 
\includegraphics[width=0.48\textwidth]{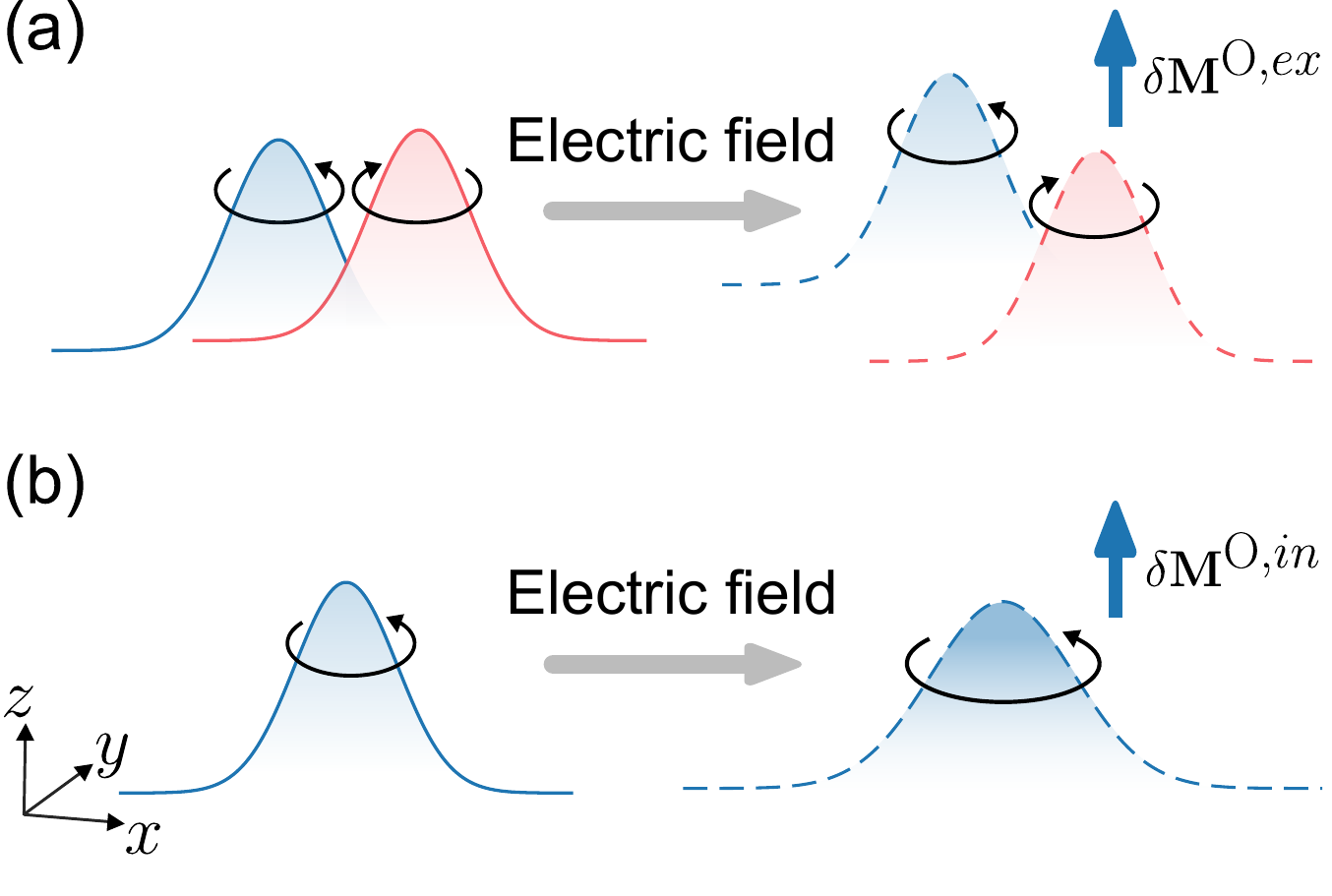}
\caption{Microscopic mechanisms of kinetic orbital magnetization. (a) Extrinsic kinetic orbital magnetization $\delta\mathbf{M}^{\text{O},ex}$ arising from a population imbalance. (b) Intrinsic kinetic orbital magnetization $\delta\mathbf{M}^{\text{O},in}$ arising from the deformation of the wave packet.}
\label{Fig: kinetic}
\end{figure}

\subsection{Orbital contribution}
Unlike the kinetic spin magnetization, the kinetic orbital magnetization can emerge via a route completely free of spin-orbit coupling~\cite{Yoda15sr, Moore16prl, Pesin17prb, Murakami20prb, Johansson24jpcm}, because the orbital motion of electrons itself carries magnetic moments. This establishes the foundation for electrically manipulating and controlling the orbital degree of freedom (i.e., orbitronics)~\cite{ZhangSC05prl, Yuriy21epl, WangP24aem, WuXS24prl, YangSY24prl, Jo24npj}.

In time-reversal-symmetric systems, the kinetic orbital magnetization can be obtained in direct analogy to $\delta\mathbf{M}^{\text{S},ex}$ [Eq.~\eqref{Eq: delta_Ms_ex}] as
\begin{equation}\label{Eq: delta_Mo_ex}
\delta\mathbf{M}^{\text{O},ex}=\frac{e\tau}{\mathcal{V}}\sum_{n,\mathbf{k}}\mathbf{m}_n \mathbf{v}_n\cdot\mathbf{E}f_0',
\end{equation}
where the Fermi-Dirac distribution function $f_0$ in Eq.~\eqref{Eq: M_orbi} is replaced by the nonequilibrium correction $f_1$, which is also referred to as the orbital Edelstein effect~\cite{Johansson24jpcm}. This extrinsic kinetic orbital magnetization can be understood as a population imbalance in momentum space [Fig.~\ref{Fig: kinetic}(a)]. Specifically, the orbital magnetic moments of time-reversal-related wave packets cancel out in equilibrium, while an electric field can shift the Fermi surface, thereby generating a finite kinetic orbital magnetization [Eq.~\eqref{Eq: delta_Mo_ex}]. We emphasize that the extrinsic kinetic orbital magnetization discussed here is restricted to the relaxation-time approximation. More intricate scattering mechanisms may provide additional extrinsic contributions to the  kinetic orbital magnetization, warranting to be further explored.

When time-reversal symmetry is further broken, there emerges a purely intrinsic kinetic orbital magnetization, analogous to $\delta\mathbf{M}^{\text{S},in}$ [Eq.~\eqref{Eq: delta_Ms_in}]. Wave packet dynamics~\cite{GaoY14prl, GaoY15prb, GaoY19fop, XiaoC21prb1, XiaoC21prb2} has revealed this kinetic magnetization as
\begin{equation}\label{Eq: delta_Mo_in}
\delta\mathbf{M}^{\text{O},in}=\frac{e}{\mathcal{V}}\sum_{n,\mathbf{k}}\mathbf{F}_n\cdot\mathbf{E}f_0.
\end{equation}
In Eq.~\eqref{Eq: delta_Mo_in}, $\mathbf{F}_n$ is the magnetoelectric tensor with its entries being 
\begin{equation}
F_n^{ij}=-2\text{Re} \sum_{m\neq n} \frac{m_{nm}^i\mathcal{A}_{mn}^j}{\varepsilon_n-\varepsilon_m}
-\frac{e}{2\hbar}\epsilon_{ilr}\partial_lg_n^{rj},
\end{equation}
where ${\mathbfcal{A}_{mn}}=\bra{m}i\partial_\mathbf{k}\ket{n}$ is the interband Berry connection, $\mathbf{m}_{m n}=\frac{e}{2}\sum_{l\neq n}(\mathbf{v}_{ml}+\delta_{lm}\mathbf{v}_n)\times {\mathbfcal{A}}_{ln}$ is the interband orbital magnetic moment,  and $\mathbf{g}_n=\text{Re}\sum_{m \neq n} {\mathbfcal{A}}_{nm} {\mathbfcal{A}}_{mn}$ is the quantum metric tensor. Besides wave packet dynamics, this intrinsic orbital kinetic orbital magnetization has also been explored by using the response theory~\cite{Yanase18prb,Koki23prb}. Physically, the intrinsic contribution arises from the distortion of the wave packet itself [Fig.~\ref{Fig: kinetic}(b)]. To be specific, the electric field directly modifies the self-rotation of the wave packet, thereby generating a finite kinetic orbital magnetization that is intrinsic to the band structure [Eq.~\eqref{Eq: delta_Mo_in}].

The kinetic orbital magnetization from the Edelstein effect has been experimentally observed in monolayer MoS$_{2}$~\cite{Lee19prl} and twisted bilayer graphene~\cite{Gorden19science}, while its intrinsic counterpart requires a further experimental exploration. Parallel to the spin case, the kinetic orbital magnetization typically coexists with the orbital accumulation driven by the orbital Hall effect~\cite{ZhangSC05prl2, Inoue08prb, Inoue08prl, Choi23nature, Dimi24apx,Dimi25npj, Dimi25prl}. A recent work~\cite{Dimi24prl} also has highlighted the importance of the extrinsic scattering mechanisms in the orbital Hall effect. Therefore, disentangling the bulk orbital magnetization from the boundary contribution of the orbital Hall effect remains a significant challenge.

\section{Challenges and outlook}
Magnetization is one of the most fundamental phenomena in condensed matter physics, yet its complete microscopic understanding has remained a long-standing challenge. Quantum geometry offers a paradigm shift, providing profound insights by revealing equilibrium and kinetic magnetization as a direct manifestation of the  geometric structure of the electronic Hilbert space.

First, despite its conceptual elegance, the current quantum geometric framework is essentially a single-particle picture. In contrast, magnetization in real materials is often deeply associated with local electron interactions and correlation effects. Establishing a deeper connection between quantum geometry and correlated electron systems, or developing a geometric framework capable of addressing genuine many-body effects, represents an exciting frontier for future exploration.

Second, within the current quantum geometric framework, spin and orbital magnetization are characterized by distinct quantum geometric quantities. However, experimentally disentangling their individual contributions to the total magnetization remains highly nontrivial. The lack of reliable experimental platforms for this \emph{spin-orbital separation} poses a major obstacle for direct quantitative comparison with theoretical predictions.

Third, extending the concept of kinetic magnetization into the nonlinear regime could be particularly promising. In this regime, magnetization scales quadratically with the applied electric field (i.e., $\delta\mathbf{M}\propto\mathbf{EE}$), enabling finite responses even in centrosymmetric systems where linear kinetic magnetization is strictly forbidden. This can significantly broaden the range of potential material platforms harboring kinetic magnetization. Recent theoretical advances have predicted nonlinear kinetic spin magnetization governed by specific quantum-geometric quantities~\cite{YangSY22prl,YangSY23prl,WangCM25prb}, while the understanding of nonlinear kinetic orbital magnetization remains largely unexplored. In contrast, experimental investigations of nonlinear kinetic magnetization are still scarce, with recent progress reported in WTe$_2$~\cite{LiaoZM24prb}.

In summary,  the quantum geometric perspective provides a fundamental understanding of magnetic phenomena by revealing that magnetization emerges naturally from the geometric structure of quantum states. We anticipate that the theoretical advances and experimental developments discussed in this Perspective will establish a solid foundation for future explorations and practical applications in spintronics and orbitronics.

\begin{acknowledgments}
This work is supported by the National Key R\&D Program of China (2022YFA1403700), Innovation Program for Quantum Science and Technology (2021ZD0302400), the National Natural Science Foundation of China (12304196, 12574173, 12350402 and 12525401), Guangdong Basic and Applied Basic Research Foundation (2022A1515111034, 2023B0303000011), Guangdong Provincial Quantum Science Strategic Initiative (GDZX2201001 and GDZX2401001), the Science, Technology and Innovation Commission of Shenzhen Municipality (ZDSYS20190902092905285), High-level Special Funds (G03050K004), and the New Cornerstone Science Foundation through the XPLORER PRIZE.
\end{acknowledgments}

\bibliography{ref}
\end{document}